\documentclass[epj]{svjour}
\usepackage{graphics}
\usepackage{hyperref}
\usepackage{lineno}
\usepackage{cite}
\usepackage{textgreek}
\begin{document}
%
\title{On-line computing challenges: detector \& readout requirements}

\author{Richard Brenner\inst{1} \and Christos Leonidopoulos\inst{2}
}                     
\offprints{}          
\institute{Uppsala University \and University of Edinburgh}
\date{Received: date / Revised version: date}
%
\abstract{
The operation at the Z-pole of the FCC-ee machine will deliver the highest possible instantaneous luminosities with the goal of collecting the largest Z boson datasets (Tera-Z), and enable a programme of Standard Model physics studies with unprecedented precision. The data acquisition and trigger systems of the FCC-ee experiments must be designed to be as unbiased and robust as possible, with the goal of containing the systematic uncertainties associated with these datasets at the smallest possible level, in order to not compromise the extremely small statistical uncertainties. In designing these experiments, we are confronted by questions on detector readout speeds with an extremely tight material and power budget, trigger systems with a first hardware level or implemented exclusively on software, impact of background sources on event sizes, ultimate precision luminosity monitoring (to the $10^{-5} - 10^{-4}$ level), and sensitivity to a broad range of non-conventional exotic signatures, such as long-lived non-relativistic particles. We will review the various challenges on online selection for the most demanding Tera-Z running scenario and the constraints they pose on the design of FCC-ee detectors. 
\PACS{
      {PACS-key}{describing text of that key}   \and
      {PACS-key}{describing text of that key}
     } 
} 
\maketitle
\section{Introduction}
The FCC-ee machine is expected to deliver the highest instantaneous luminosities ever achieved, forcing a re-evaluation of the requirements for trigger and data acquisition (DAQ) systems.

The conventional wisdom is that the trigger systems of FCC-ee experiments must rely on simple (low- or minimum-bias\footnote{In the trigger context, minimum-bias events are inelastic events, typically from both diffractive and non-diffractive processes, which have been selected with as little bias as possible during the online filtering carried out by a trigger system. A low-bias trigger typically applies an additional requirement, like the existence of a low-energy particle candidate such as an electron, muon or jet, in the event.}) triggers with built-in redundancy, e.g. calorimeter-based, muon-based or tracker-based. For example, in the LEP era \cite{lep}, the online selection was established from calorimeter- and tracker-based triggers. For the ILC studies \cite{ilc}, the assumption has been that the experiments will rely on a `triggerless' DAQ (i.e. no first-level hardware trigger), exploiting the relatively small collision rates. It is worth mentioning that LHCb~\cite{LHCbTDR}, one of the current experiments, is going to collect all detector data from collisions and feed it into an event selector that will run entirely in software. The experimental environment at FCC-ee is, however, very different from that at LHCb. The event rate is significantly lower than at a hadron collider, but the material budget is much tighter which limits the services and readout bandwidth. Compared with previous experiments at lepton colliders,  the  challenge for FCC-ee experiments is  the very large data rates ($\sim$ 200 kHz when running at the Z-pole), which are orders of magnitude larger than at LEP and are significantly higher than at Belle II.

In this essay, we review studies of hardware and software solutions that will allow FCC-ee experiments to record all of the interesting physics events with very high efficiency and redundancy, leading to minimum uncertainties and biases in the experimental measurements.

\section{Event sizes and collision rates at FCC-ee}
The parameters of interest in a trigger \& DAQ (TDAQ) design are:
\begin{itemize}
\item the rate of interesting physics events to record
\item the rate of irreducible background events
\item the average event size
\item the data throughput, which is the product of event-size and the interaction rate, corresponding to the data volume per unit time. The data throughput can be derived from the previous three parameters, and is the key element of the TDAQ that determines the read-out and write-out capacity of the system. 
\end{itemize}

At FCC-ee, the nominal instantaneous luminosities  per interaction point range from a maximum of $230\times10^{34}\,{\rm cm}^{-2}\,{\rm s}^{-1}$ at the Z-pole, down to 28 (WW), 8.5 (ZH) and 1.8 $\times 10^{34}~{\rm cm}^{-2}\,{\rm s}^{-1}$ ($\mathrm{t\bar{t}}$) for the other running scenarios \cite{cdr}. In the context of TDAQ design, we concentrate on the most rate-demanding operations at the Z-pole. The total event rate in this scenario is around 200\,kHz (Table~\ref{tab:evtrate}), with the dominant processes including Z-production (100\,kHz), Bhabha scattering (50\,kHz), \textgamma \textgamma $\rightarrow$ hadron production (30\,kHz) and beam background (20\,kHz) \cite{eventsize}. 

The basic assumption for trigger operation of the  FCC-ee experiments is that all interesting physics can be stored for offline analysis with $\sim$ 100\% efficiency, and that the beam background (expected to be $\sim$10\% of the total event rate at the Z-pole) is not a major consideration for DAQ. To evaluate average event sizes for each process and their overall contribution to data throughput, several assumptions are made about detector granularities, and that the experiments will be relying on stable zero-suppression\footnote{Zero-suppression is the process during which data in the front-end electronics or the readout system is only recorded above a certain significance level.}  during DAQ. The studies described in Ref.~\cite{cdr} have been performed using early prototypes of the {\tt CLD}~\cite{cld} and {\tt IDEA}~\cite{idea} detectors and are summarised in Table~\ref{tab:evtsize}. The main conclusion is that with an appropriate zero-suppression scheme, the major contribution to the average event size for the {\tt IDEA} detector is from physics, and it should be possible to keep the main backgrounds (e.g. synchrotron radiation) under control at a relatively small fraction of the total event rate. 
\begin{table}
\centering
\caption{Event rates expected for various processes at the Z-pole at the FCC-ee \cite{eventsize, cdr}. The beam background is expected to be $\sim$10\% of the total event rate.}
\label{tab:evtrate}       
\begin{tabular}{lr}
\hline\noalign{\smallskip}
Physics process & Rate (kHz) \\
\noalign{\smallskip}\hline\noalign{\smallskip}
Z decays & ~100  \\
$\gamma \gamma \rightarrow$ hadrons & ~~~30 \\
Bhabha & ~~~50 \\
Beam background & ~~~20 \\
\hline
Total & $\sim$ 200 \\
\noalign{\smallskip}\hline
\end{tabular}
\end{table}
\begin{table}
\centering
\caption{Average event data rates expected for the {\tt CLD} and {\tt IDEA} subdetectors at the Z-pole for the FCC-ee \cite{eventsize, cdr}. }
\label{tab:evtsize}       
\begin{tabular}{lcc}
\hline\noalign{\smallskip}
Subdetector & Physics & Background/noise \\
\noalign{\smallskip}\hline\noalign{\smallskip}
CLD Vertex Detector & 150 MB/s & 6 GB/s \\
CLD Tracker & 160 MB/s & 10 GB/s \\
IDEA Drift Chamber & 60 GB/s & 2 GB/s \\
IDEA Si Wrapper & 32 MB/s & 0.5 GB/s \\
IDEA DR Calorimeter & 10 GB/s & 1.6 TB/s $^*$ \\
IDEA pre-shower & 320 MB/s & 820 MB/s \\
IDEA Muon Detector & 4 MB/s & 67 MB/s \\
\noalign{\smallskip}\hline
\multicolumn{3}{l}{$^*$ Assuming no suppression for isolated counts}
\end{tabular}
\end{table}

One of the most important ingredients of precision measurements at lepton colliders is the accurate determination of the integrated luminosity. The FCC-ee programme has the very ambitious goal of measuring the absolute luminosity at the $10^{-4}$ level, with the relative luminosity values between energy-scan points measured 10 times better, at the $10^{-5}$ level. The challenges for the FCC-ee luminosity monitor design and possible solutions are discussed in a separate essay \cite{lumi}. 

\section{Triggerless design}
A software-based online selection (the so-called `triggerless' design) provides a flexibility that cannot be matched by traditional first-level hardware-based filtering systems. Typically, as luminosity increases, a calorimeter- and muon-based filtering system cannot sustain the trigger rates without an increase of threshold values. This means that a trigger inefficiency, which cannot be compensated by a higher integrated luminosity, is incurred. A purely software trigger also allows increased complexity in the online selection by combining information from different systems, provided the algorithmic execution fits within the processing time budget.

For an FCC-ee experiment, the major challenge to a triggerless design is the very high instantaneous luminosity, in particular at the Z-pole ($2.3 \times 10^{36}~{\rm cm}^{-2}\,{\rm s}^{-1}$). R\&D studies assume that zero-suppression will be routinely applied at read-out time. However, this requirement necessitates not only a careful calibration (and alignment) of the subdetector systems, but also a technical solution for a trigger that can be deployed online, and updated in semi-real time. In that sense, smooth and stable running conditions and a robust monitoring system are of paramount importance, as has been demonstrated in operation at LEP and the B factories. 

Detector choices can, of course, have an impact on the trigger and DAQ design. Consider, for example, the following options:
\begin{itemize}
\item Tracking: a Time Projection Chamber (TPC) which is favoured by tracking experts for its lightweight design, cannot be read out every 20\,ns. A TPC-based detector would, therefore, necessitate a hardware-based first filtering level in order to reduce the readout of the TPC.
\item Calorimetry: a fine-granularity but noisy calorimeter, where the application of zero-suppression at the trigger may not be straightforward. For example, the noise data rates for the IDEA DR Calorimeter (Table \ref{tab:evtsize}) where the working assumption is no suppression for isolated counts. The noise in the Z-pole running scenario contributes significantly more to the average event data rates (1.6 TB/s) than the physics (10 GB/s), a constraint that would interfere with the optimisation of trigger efficiency of electromagnetic showers.
\end{itemize}
It is therefore necessary to balance the detector requirements against operational considerations and constraints on the trigger and DAQ systems when designing the FCC-ee experiments. 
\vspace{3mm}

With TDAQ technology evolving rapidly, and FCC-ee still far into the future, it is perhaps too early to discuss details of concrete TDAQ designs and implementation. However, it is still instructive to  review some recent advances in HEP experiments that may be relevant when designing TDAQ systems for FCC-ee experiments. We do this in the following sub-sections, by reviewing the LHCb TDAQ system, the readout of ultra-light vertex detectors, and tracking with ultra-light TPCs.

\subsection{A recent example: the LHCb TDAQ system}

The LHCb experiment has opted for a triggerless design for the Run-3 upgrade \cite{LHCbTDR, LHCb:2021kxm, lhcb}. Event-selection is implemented entirely in software in order to maximise physics yield, increase the amount of data collected, minimise cost and retain flexibility. 

LHCb employs FPGAs in the `middle-layer' of the DAQ, which is the electronics bridging the readout of the detectors to the optical fibres that bring the signal to the data centre on the surface. The middle-layer is physically located at the periphery of the detector where the radiation level is moderate, and this allows LHCb to operate FPGAs near the detector hardware. The experiment applies zero-suppression directly at the detector readout. FPGAs will be tested up to a total ionising dose of 200 krad and a high energy hadron (HEH) fluence of $1.2 \times 10^{12}\, {\rm cm}^{-2}$ \cite{fpga}. Given the reduced geometrical acceptance of the detector, the total event-size is comparatively small for an LHC experiment and yields $\sim$ 100 kBytes. LHCb is able to perform online selection with offline-like reconstruction, eliminating a significant source of biases.

There are two levels of software filtering: the first one, running on GPUs, reduces the data throughput from 32\,Tb/s to 1-2\,Tb/s, with the second one bringing it down further to 80 Gb/s. A large storage buffer allows the effective decoupling of the two sub-systems. Calibration and alignment are performed in `semi-live' mode, while the data is being buffered. The challenges to the LHCb TDAQ design are the large memory consumption and the capacity of the network (which needs to collect data in a single physical location from 478 FPGA boards  for event selection). High data rates in a fixed bandwidth system can lead to network congestion. To avoid any data loss, LHCb requires a lossless network. The approach followed is to prioritise the network traffic by application, in a technique known as `traffic-shaping' that helps optimise performance and improve the overall latency of the data flow. 

The biggest worry that comes with the LHCb design is the scalability and reliability of the network. TDAQ experts' estimates are that the online system can sustain up to a factor of three increase in network traffic without any substantial changes in its architecture \cite{lhcb}. This scalability is expected to be adequate for the duration of Run-3. 

\subsection{Readout of ultra-light vertex detectors}

The development of Monolithic Active Pixel Sensors (MAPS) has been pioneered by the communities developing detectors for experiments with lepton and heavy ion beams, where low material budget is essential.  The most advanced tracker in operation that is using MAPS technology is the Inner Tracking System of ALICE \cite{alice}.  It uses a custom made ALPIDE ASIC \cite{alpide}, which has a thick high-resistivity epitaxial layer for sensing charged particles with readout electronics embedded in a low resistivity n-well. The pixels in the tracker are 29\,\textmu m  x 27\,\textmu m in size.  Every pixel is equipped with an amplifier, comparator and a 3-hit deep register for storing data. Readout of the 1024 $\times$ 512 pixel matrix is done with electronics on the periphery that allow up to 1200 Mbps data rate, giving the chip a theoretical maximum hit data transfer capacity of 6 MHz/cm$^2$.  The ASIC has a continuously active front-end and the readout is zero suppressed, offering both triggered and continuous readout with time-stamped hit information. The chip has  demonstrated a readout rate of 100 kHz for Pb-Pb interactions. The performance of ALPIDE already meets the rates presented in Table \ref{tab:evtsize}. The material budget of ALPIDE is around 0.3$\%~ X_0$/layer, which is higher than the target of 0.15$-$0.2$\% ~X_0$/layer for a vertex detector in a future $\mathrm{e^+e^-}$ experiment. Development will be required to optimise the readout performance with the material and power budgets available. 

There are many improvements to the MAPS technology that can be expected for the FCC. The charge collection in the pixel will become significantly faster with HV-CMOS technology, which is based entirely on drifting. Improved time-stamping of hits will allow full 4D tracking. Having 4D tracking capability at FCC-ee detectors opens up the possibility of separating multiple interaction points, and could improve the sensitivity to, for example, slowly propagating exotic particles.

The tracker community is continuously developing ways to reduce the amount of passive material in the tracker. Low power electronics combined with evaporative CO$_2$ cooling with microchannels etched in silicon are the most recent developments to reach operation in an experiment.  Wireless data transmission is being developed by the WADAPT collaboration~\cite{WADAPT} with the goal of demonstrating the feasibility of this technology for low power,  Gbps rate, and the short-range data links employed in tracking detectors. The technology will increase the readout bandwidth without increasing the amount of material in the detector. Local wireless communication inside the tracker can also be used for data reduction inside the tracker prior to readout.   

\subsection{Tracking with ultra-light TPC}

Time projection chambers are an attractive option for lepton collider experiments since low-mass trackers offer high hit precision, which in turn delivers excellent momentum resolution. As discussed earlier, an important consideration is whether their readout is compatible with the requirements of a triggerless design. Here we review methods recently developed by the ALICE collaboration to tackle this problem. 

The challenge in developing a TPC with triggerless readout for ALICE is to master the huge out-of-bunch pile-up during TPC drift time. This is particularly important for searches of rare signals in a massive background. The requirement is to store full minimum bias samples for analysis. The readout rate for a conventional TPC is limited to  about 3\,kHz by the gating grid that blocks the back flow of ions.  Replacing the multi wire proportional chamber with a gas electron multiplier removes the need for a gating grid since the intrinsic back flow is low. It also opens the possibility for triggerless readout since the TPC can now be operated in continuous mode.  The number of ions entering the TPC region is still large enough to produce an electric field that distorts the path of the electrons during drifting. The effect of the distortion is rate dependent and must be corrected in order to maintain the intrinsic resolution of the TPC. The correction of space charge distortion will be an important part of the TDAQ.  

Two main methods have been developed for correcting the space charge distortions in ALICE:

\begin{itemize}
    \item Track-based, where tracks are refitted using information from all tracking detectors.  It is very accurate but requires large samples of tracks to deliver the necessary statistical precision.
    \item Integration of currents arriving on the TPC end plates. It requires a full ion history which, in turn, requires continuous readout. The method is probably less accurate than the track-based method. 
\end{itemize}

In ALICE, the processing of TPC hits is based on 10$-$20\,ms time frames. The calibration intervals need to be fast ($\sim$ 5\,ms) to be able to correct for short time fluctuations.  
 
Another challenge is the absence of absolute $z$-coordinates together with the aforementioned space charge distortions. To fully take advantage of the  high intrinsic precision of the TPC, the TDAQ will need to have the resources to carry out track reconstruction to calibrate and compensate for any distortions.   

Groups developing the new TPC for ALICE have been actively involved in technological innovations for $\mathrm{e^+e^-}$ colliders through the AIDA programme~\cite{aida}. The improvements introduced in the TPC make it a serious alternative for FCC-ee experiments because of the low material budget and continuous tracking capability. The interplay of the TPC and MAPS based tracking needs to be optimised to have the best physics performance.

\section{The challenge of long-lived non-relativistic particles}

The inclusion of the tracking detectors in the fast data handling path will continue to be a big challenge for TDAQ in current and future experiments. The large quantities of data generated by tracking detectors need to be rapidly transferred to off-detector processors. This is not easily done due to the strict requirement of a low material budget for the tracking detectors resulting from the physics goal of high precision measurements. The availability of tracking information through the full TDAQ processing chain  will be important for expanding the physics programme to new exotic physics signatures that may otherwise have escaped undiscovered. 

Many new models beyond the Standard Model (SM) postulate Long Lived Particles (LLP) that have a lifetime long enough for them to travel some measurable distance inside the detector before decaying. These exotic particles differ from the known long lived particles in the SM in terms of their unusual ionisation and propagation properties. The physics programme at FCC will require reconstruction algorithms in searches for LLP to capture these signatures.
Searches for new physics with LLP is a challenge for TDAQ due to the various signatures which have appearing and disappearing tracks that do not point to the primary vertex. The capability to select LLP events in real-time data is usually not a priority in the design phase of an experiment because they are not the primary physics motivation. The complexity of the signature also makes it difficult to find good metrics for the production of design requirements for LLP physics. 

A lesson from current experiments which should be considered in future designs is that it is important to have a clear strategy for LLP at an early stage and this may require additional (and potentially distant) detectors to be integrated in the TDAQ system. If the readout has to be included in the trigger, the experiment needs a readout segmentation that lends itself to LLP triggering. This can include hardware track triggering being instrumented in the tracker. The LLP physics programme  would, however, benefit from a triggerless readout including timing information of every hit. This would allow studies of out-of-bunch and out-of-time particles. 

\section{Machine Learning}
In recent years we have witnessed an explosive growth of machine learning techniques in HEP applications. Indeed, there is a strong R\&D activity concerning the deployment of machine learning technologies in both off-the-shelf commercial processors and FPGAs with a more limited computing footprint \cite{NNFPGA, GNN}. Some examples include: front-end data compression, particle identification with multivariate classifiers, pattern recognition, tracking and reconstruction with neural networks and regression for improved resolution. It is widely anticipated that some of these innovations will soon find their way into TDAQ systems\cite{ATLASTDAQ, CMSTDR}. We expect that in the next five years these developments will mature and form the basis for proto-TDAQ design contributions to the FCC feasibility study. 

\section{Conclusions}
The biggest challenge for FCC-ee experiments is the all-time record high instantaneous luminosity at the Z-pole. TDAQ systems sustaining similar data throughput rates will already be operational at the LHC. The baseline assumption for future lepton collider experiments is that they will rely on a software-only triggering system with $\sim$100\% efficiency and built-in redundancy. The possibility of using an ultra-light tracking detector, such as TPC, in a triggerless system requires the management of the very large out-of-bunch pile-up and its operation in continuous mode. Full timing information of detector hits is expected to be fully exploited in the TDAQ of FCC-ee experiments because it will be needed for calibration, reconstruction and in searches for new exotic particles. Of course, when making projections on rapidly developing technologies and how these may affect the design of future experiments, it should be remembered that they come with a degree of uncertainty.   
%
%
%

%
%

\end{document}